\documentclass[iop,apj]{emulateapj}

\shortauthors{Ruan, Anderson, Dexter et al.}
\shorttitle{Temperature Fluctuations in Accretion Disks}
\begin{document}
\title{Evidence for Large Temperature Fluctuations in Quasar Accretion Disks From Spectral Variability}
\author{John~J.~Ruan\altaffilmark{1,2}, 
  Scott~F.~Anderson\altaffilmark{2}, 
  Jason~Dexter\altaffilmark{3},
  Eric~Agol\altaffilmark{2}
  }
\altaffiltext{1}{ Corresponding author: jruan@astro.washington.edu} 
\altaffiltext{2}{Department of Astronomy, University of
Washington, Box 351580, Seattle, WA 98195, USA}
\altaffiltext{3}{Departments of Physics and Astronomy, University of 
California, Berkeley, CA 94720, USA}

\keywords{quasars: general}

\begin{abstract}

The well-known bluer-when-brighter trend observed in quasar variability is a signature
of the complex processes in the accretion disk, and can be a probe of the quasar
variability mechanism. Using a sample of 604 variable quasars with repeat spectra in 
SDSS-I/II, we construct difference spectra to investigate the physical causes of this 
bluer-when-brighter trend. The continuum of our composite difference spectrum is
well-fit by a power-law, with a spectral index in excellent agreement with previous results. 
We measure the spectral variability relative to the underlying spectra of the quasars, which 
is independent of any extinction, and compare to model predictions. We show that our
SDSS spectral variability results cannot be produced by global accretion rate fluctuations 
in a thin disk alone. However, we find that a simple model of a inhomogeneous disk with
localized temperature fluctuations will produce power-law spectral variability over optical 
wavelengths. We show that the inhomogeneous disk will provide good fits to our observed
spectral variability if the disk has large temperature fluctuations in many independently 
varying zones, in excellent agreement with independent constraints from quasar microlensing 
disk sizes, their strong UV spectral continuum, and single-band variability amplitudes. 
Our results provide an independent constraint on quasar variability models, and 
add to the mounting evidence that quasar accretion disks have large localized temperature 
fluctuations.
\end{abstract}

\section{Introduction}

	A well-known characteristic of the quasar phenomena is 
their strong flux variability in many wavelength regimes, including the radio, optical, X-ray, and 
$\gamma$-rays \citep{ul07}. In particular, the rise of optical large-scale time-domain imaging
surveys has led to many recent investigations of broadband quasar optical variability 
properties using large numbers of well-sampled light curves, especially for use in quasar 
selection \citep{ke09, ko10, ma10, schmidt10, bu11, ki11, ma11, ru12, an13, zu13}. These studies have 
generally revealed that quasars are stochastically variable on the $\sim$10-20\% level in flux on 
long time-scales, and show weaker, correlated variability on timescales $\lesssim$ 1 year in the 
rest-frame. The physical cause of quasar variability is still unclear, but since the optical continuum 
is likely to be dominated by emission from the accretion disk, some studies have 
suggested that changes in the global accretion rate in the disk may be able to produce 
such effects \citep{pe06, li08, zu12}. These claims appear to be supported
(although not implied) by various observed trends between optical variability amplitude,
black hole mass, and luminosity in different quasars \citep{ho94, ga99,va04, wi08, ba09, 
ke09, ma10, zu12}, in turn suggesting that the differences in variability across a sample of quasars 
may be driven at least in part by their Eddington ratio.

	Accretion rate fluctuations are expected in individual quasars due to processes in the disk 
such as the magnetorotational instability \citep[MRI,][]{ba91}, which is now generally accepted to 
operate in a wide range of accretion flows. Results from non-radiative global simulations of
thin magnetized disks have also shown that such accretion flows are almost certain to be 
highly turbulent \citep[e.g.][]{am01, ar03, no09a, no09b, pe10}, but the exact 
characteristics of a MRI flow in a global radiative MHD simulation are currently unclear,
especially their stability in the radiation pressure-dominated regime 
\citep[see discussions in][]{hi09, ja12, ji13}. Localized temperature fluctuations in highly 
turbulent disks will also cause flux variations, and such a scenario may be expected since quasar 
accretion disks are too large to vary coherently in flux over the short variability timescales observed. 
The characteristic timescales of quasar flux variability have also been shown to be consistent with
the thermal timescale \citep{ke09}, independently motivating accretion disk models involving localized temperature fluctuations. Furthermore, it has been shown that such models of inhomogeneous accretion 
disks can also simultaneously explain quasar microlensing disk sizes, their strong UV spectral continuum, and 
single-band optical variability properties \citep{de11}. 

	Although the relative roles of global accretion rate fluctuations and localized temperature 
fluctuations in accounting for the observed flux variability are unclear, an 
additional probe is provided by the characteristic bluer-when-brighter trend observed in 
studies of quasar spectral variability \citep{cu85, gi99, tr01, tr02, va04, wi05, me11, sa11, sc12, me13}. 
This trend is almost certainly a direct consequence of the underlying quasar variability 
mechanism, and thus provides an independent test of quasar variability models. Indeed, 
both global accretion rate changes and localized temperature fluctuations in accretion disks 
will generically produce bluer-when-brighter trends, but the details of the predicted 
trend are dependent on the details of the model. Intriguingly, previous investigations of
quasar spectral variability have sometimes resulted in disparate conclusions.

	\citet{wi05} used a sample of 315 pairs of repeat spectra of variable quasars from the 
Sloan Digital Sky Survey \citep[SDSS,][]{yo00}, and constructed `difference spectra', by taking 
the spectrum of each quasar at the higher-flux epoch and subtracting the spectrum at the 
lower-flux epoch; this effectively isolates the variable part of the spectrum. After applying a 
wavelength-dependent spectrophotometric recalibration on each pair of repeat spectra, they 
find that the resulting composite quasar difference spectrum has a steeper power-law index than 
the composite of the individual spectra, showing that quasars are indeed bluer-when-brighter. 
Based on the composite difference spectrum from that study, \citet{pe06} fitted synthetic 
difference spectra generated from a simple Shakura-Sunayev thin disk model \citep{sh73}, 
and showed that the composite difference spectrum can be produced from a simple thin disk in 
which the global accretion rate has changed \citep[also see][]{sa11}.

	In contrast, \citet{sc12} used a sample of 9093 multi-band quasar light curves from SDSS Stripe 
82 to study the bluer-when-brighter trend using many epochs of broadband photometry. After
correcting for the effects of broad emission lines (which are well-known to be less variable than 
the continuum) in each filter, they compare their results to spectral variability predictions 
from accretion rate fluctuations in a simple thin disk, as well as more detailed static disk 
models. They find that accretion rate changes in these disk models \emph{cannot} 
reproduce the strong bluer-when-brighter trend, and instead suggest that ephemeral hot spots 
on the accretion disk may be needed \citep[also see][]{tr02, me13}.

	The conclusions of \citet[][hereafter WI05]{wi05} and \citet[][hereafter SC12]{sc12} appear 
at first glance to be at odds: can the bluer-when-brighter trend observed in quasar be explained
by fluctuations in the global accretion rate in a simple thin disk, or are localized temperature 
fluctuations needed? In this paper, we revisit the spectral variability study of WI05 using a larger 
sample of repeat quasar spectra culled from the full SDSS-I/II data set, to investigate these 
apparently discordant results. We will also compare our spectral variability results to the 
recently-developed time-dependent model of inhomogeneous accretion disks by \citet{de11}. 
The structure of this paper is as follows: In Section 2, we describe the construction of the sample 
of repeat quasar spectra used in this study, and our spectrophotometric recalibration. In Section 3, 
we discuss the properties of the quasars' difference spectra, as well as the construction of composite 
spectra and composite difference spectra. In Section 4, we compare our results to the previous 
studies of WI05 and SC12 using global accretion rate fluctuations in a thin disk. In Section 5, we 
discuss the quasar spectral variability predicted from a time-dependent model with temperature 
fluctuations, and show that our observations are well-fit by such inhomogeneous disks. In Section 6, 
we discuss the connection between disk properties and the resultant spectral variability, as well as 
other variability mechanisms that might match the observations. We summarize and conclude in 
Section 7. 

\section{Data Selection and Reduction}
\subsection{SDSS-I/II Repeat Spectra}
	All spectroscopic data used in our paper is from the SDSS-I/II, which is publicly available 
in its entirety as part of SDSS Data Release 7 \citep[DR7,][]{ab09}. The SDSS-I/II obtained 
follow-up spectra of  approximately $1.6\times10^6$ objects, including more than 
$1.1\times10^5$ quasars \citep{schneider10}, primarily selected by optical color from the imaging 
portion of the survey \citep{ri02}. The two fiber-fed SDSS spectrographs utilize a
total of 640 fibers plugged into holes drilled onto plates, which are placed at the telescope 
focal plane. During the normal course of operations, multiple 15-minute exposures of each plate 
are taken, and spectra from exposures within approximately a month are typically coadded together. 
The spectral reduction and calibration using the SDSS Spectro2d pipeline are described in \citet{st02}.
Occasionally, entire plates may be reobserved and coadded separately as a second epoch of 
spectra. This may occur if the first epoch did not reach sufficiently high signal-to-noise ratios (SNR), or 
in some cases by design as part of the survey plan. For these multiply-observed plates, no attempt 
was made to ensure that the same fiber was plugged into the same hole on the plate, and so 
spectra of the same object may have different fiber numbers in the different epochs, even though 
the plate number is identical. For more details about these repeatedly observed plates, we refer to 
discussions in \citet{wi05}.

	Although multiple epochs of SDSS spectra are also possible due to spatial overlaps in the sky
between adjacent plates, \citet{va04} showed that additional calibrations based on non-variable
stars on the same plates enhance sensitivity to the wavelength-dependent variability properties 
of quasars. Overlapping regions on adjacent plates are generally small, and will not have many
non-variable stars in the overlapping regions to accurately recalibrate the quasar spectra in the
same regions. Thus, we focus only on multi-epoch spectra from plates that have been reobserved 
in their entirety, which ensures that multi-epoch spectra of many calibration stars are available 
in addition to the quasars. We also note that more epochs of spectra for many SDSS DR7 quasars 
are now publicly available as part of the SDSS-III Baryon Oscillation Spectroscopic Survey 
\citep{da13}. However, SDSS-III utilizes a newer spectrograph, fiber system, and 
spectral reduction pipeline; robust comparison of continuum properties 
of spectra between SDSS-I/II and SDSS-III are difficult and thus not considered here. 

	Using the plate list of 2880 observations of all 2698 unique plates in DR7 from SDSS-I/II, we 
select only those which are part of the main SDSS survey (and its primary reduction pipeline) by 
requiring the flag SURVEY = `sdss'. A data quality cut is then made by requiring the flag
PLATEQUALITY = `good' or `marginal'; plates which pass this quality cut have SNR $>$ 9
and less than 13\% problematic pixels. From the remaining plates, we select those which have 
multiple observations with time-lag $>$30 days between each pair of epochs in the observed frame.
This time-lag cut is physically motivated from photometric studies of quasar light curves, 
which have shown that quasars are generally not variable above $\sim$1\% in flux on such short 
timescales \citep{ke09, ko10, ma11}. We note that a few plates had three or more observations; 
in such cases, we use all unique pairs of observations of each plate that pass all the above criterion.
There are a total of 71 unique pairs of plate observations in SDSS-I/II, which are listed in 
Table 1. Figure 1 shows the distribution of the timelags for these 71 pairs of plate observations, 
which range from 30 days to about 3 years in the observed-frame.  

\begin{deluxetable}{ccccc}
\tablecolumns{5}
\tablewidth{0pt}
\tablecaption{All unique pairs of repeatedly-observed plates in SDSS-I/II with time-lag $>$30 days,  PLATEQUALITY = `good' or `marginal', along with the number of quasars and variable quasars on each plate.}
\tablehead {\colhead{Plate} \vspace{0cm}& \colhead{High-SNR} & \colhead{Low-SNR} &\colhead{Quasars} & \colhead{Variable}\\ \colhead{Number} \vspace{0cm}& \colhead{MJD} & \colhead{MJD} & & \colhead{Quasars}}    
\startdata
291 & 51660 & 51928 & 21 & 9\\
293 & 51994 & 51689 & 54 & 18\\
296 & 51578 & 51984 & 26 & 7\\
297 & 51663 & 51959 & 23 & 15\\
300 & 51666 & 51943 & 41 & 6\\
301 & 51641 & 51942 & 45 & 17\\
304 & 51957 & 51609 & 26 & 12\\
306 & 51690 & 51637 & 35 & 1\\
309 & 51666 & 51994 & 47 & 14\\
340 & 51691 & 51990 & 24 & 3\\
351 & 51780 & 51695 & 36 & 3\\
352 & 51789 & 51694 & 29 & 6\\
360 & 51780 & 51816 & 69 & 6\\
385 & 51783 & 51877 & 49 & 5\\
390 & 51816 & 51900 & 30 & 7\\
394 & 51876 & 51812 & 31 & 6\\
394 & 51812 & 51913 & 28 & 5\\
394 & 51876 & 51913 & 31 & 5\\
404 & 51877 & 51812 & 14 & 1\\
406 & 51817 & 51869 & 48 & 6\\
406 & 51876 & 51817 & 50 & 4\\
406 & 51817 & 51900 & 49 & 5\\
406 & 52238 & 51817 & 48 & 19\\
406 & 51900 & 51869 & 54 & 6\\
406 & 52238 & 51869 & 47 & 19\\
406 & 52238 & 51876 & 51 & 19\\
406 & 52238 & 51900 & 49 & 20\\
410 & 51816 & 51877 & 83 & 22\\
411 & 51873 & 51817 & 28 & 8\\
412 & 51871 & 51931 & 30 & 3\\
412 & 51871 & 52235 & 29 & 8\\
412 & 51871 & 52250 & 31 & 8\\
412 & 51871 & 52254 & 30 & 13\\
412 & 51871 & 52258 & 29 & 11\\
412 & 52235 & 51931 & 32 & 12\\
412 & 52250 & 51931 & 35 & 13\\
412 & 52254 & 51931 & 32 & 11\\
412 & 51931 & 52258 & 32 & 11\\
413 & 51821 & 51929 & 46 & 3\\
414 & 51869 & 51901 & 39 & 3\\
415 & 51879 & 51810 & 39 & 4\\
416 & 51885 & 51811 & 68 & 31\\
418 & 51884 & 51817 & 72 & 16\\
419 & 51812 & 51868 & 69 & 7\\
419 & 51812 & 51879 & 64 & 24\\
422 & 51878 & 51811 & 26 & 1\\
476 & 52027 & 52314 & 80 & 20\\
483 & 51942 & 51902 & 78 & 8\\
525 & 52029 & 52295 & 52 & 19\\
547 & 51959 & 52207 & 66 & 19\\
662 & 52178 & 52147 & 37 & 5\\
803 & 52264 & 52318 & 4 & 1\\
810 & 52326 & 52672 & 4 & 0\\
814 & 52370 & 52443 & 47 & 7\\
820 & 52405 & 52438 & 80 & 3\\
960 & 52466 & 52425 & 31 & 1\\
1028 & 52562 & 52884 & 3 & 2\\
1034 & 52525 & 52813 & 2 & 1\\
1037 & 52826 & 52878 & 1 & 0\\
1512 & 53035 & 53742 & 19 & 7\\
1670 & 53438 & 54553 & 41 & 19\\
1782 & 53383 & 53299 & 31 & 3\\
1905 & 53613 & 53706 & 21 & 5\\
1907 & 53265 & 53315 & 27 & 3\\
2009 & 53857 & 53904 & 44 & 3\\
2061 & 53405 & 53711 & 17 & 8\\
2252 & 53565 & 53613 & 0 & 0\\
2294 & 54524 & 53733 & 63 & 37\\
2394 & 54518 & 54551 & 0 & 0\\
2474 & 54333 & 54564 & 7 & 1\\
2858 & 54498 & 54464 & 1 & 1\\
\enddata
\end{deluxetable}

\begin{figure}[t]
\begin{center}
\includegraphics[width=0.49\textwidth]{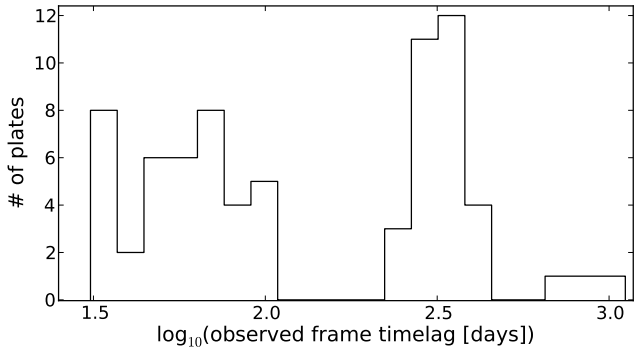} 
\caption{Distribution of the observed-frame time-lags between repeat observations of the 71 pairs of plate observations used in our sample (listed in Table 1).}
\end{center}
\end{figure}

\subsection{Spectrophotometric Recalibration}
	We perform a wavelength-dependent spectrophotometric recalibration on all pairs 
of repeat quasar spectra on each plate, by first producing a `recalibration spectrum' for 
each pair of observations based on the non-variable stars on each plate. This is 
done following WI05 with only minor modifications, to allow for faithful comparison to previous 
work. We note here that to facilitate our difference spectra analysis, we have resampled all 
spectra and their uncertainties onto a common wavelength grid of the form log$_{10}\lambda = 2.602 + 0.001a$, for integers $a$ from 0 to 1400 ($\lambda$ in \AA~units). This is approximately a factor 
of 10 coarser than the actual SDSS spectral resolution, but appropriate for our investigation 
of the continuum properties of quasars. The resampling is done using a simple linear 
interpolation, and the resulting common wavelength grid covers 400 to 10,046\AA, wide enough 
for all rest-frame spectra of the quasars in our sample. As part of the interpolation to the common 
wavelength grid, we mask out problematic pixels in the each spectrum which had SDSS pipeline 
flags set for NOPLUG, BADTRACE, BADFLAG, BADARC, MANYBADCOLUMNS, 
MANYREJECTED, NEARBADPIXEL, LOWFLAT, FULLREJECT, SCATTEREDLIGHT, NOSKY, 
BRIGHTSKY, COMBINEREJ, REDMONSTER \citep[for details on these flags, see][]{st02}. 
We consider only pairs of repeat spectra of objects for which $<$20\% of pixels are rejected in 
both epochs.

	For each pair of plate observations, all pairs of stellar spectra are 
selected by  requiring the SDSS Spectro1d pipeline classification of both spectra to have 
CLASS = `STAR' and their SUBCLASS classification to be identical between the 
two epochs. To remove the stars that have significantly 
varied between the two epochs, we integrate each pair of stellar spectra and calculate the 
relative change in flux of the star $\Delta f/f = |(f_1 - f_2)| / (0.5f_1 + 0.5f_2)$, where $f_1$ and 
$f_2$ are the integrated fluxes in the two epochs. Stars with large $\Delta f/f$ are unsuitable
for use in the spectrophotometric recalibration due to their variability. Since $\Delta f/f$ is dependent 
on the SNR of the spectra, we follow the procedure of WI05 to include the flux uncertainties on the 
spectra in the variability selection by placing a variability cut on $\Delta f/f$ as a function of the SNR; 
Figure 2 shows the distribution of $\Delta f/f$ against
the SNR of high-SNR epoch spectra for all stars on the 71 plates. We bin the stars in Figure 2 
into 13 equally-sized bins of SNR, and calculate the 90th percentile in $\Delta f/f$ in each bin. 
We then fit an envelope with these 13 points of 90th percentile $\Delta f/f$ to an 
exponential function of the form $\Delta f/f$  = 0.53exp(SNR/$-0.57)+0.10$. Stars with 
$\Delta f/f$ below this envelope are considered `non-variable' (Figure 2), and are 
used in the subsequent spectrophotometric recalibration. There were 6327 stars in total over 
the 71 plates, 5615 of which were deemed to be non-variable. On a typical plate, a median of 
47 non-variable stars were used in the spectrophotometric recalibration.

\begin{figure}[t]
\begin{center}
\includegraphics[width=0.49\textwidth]{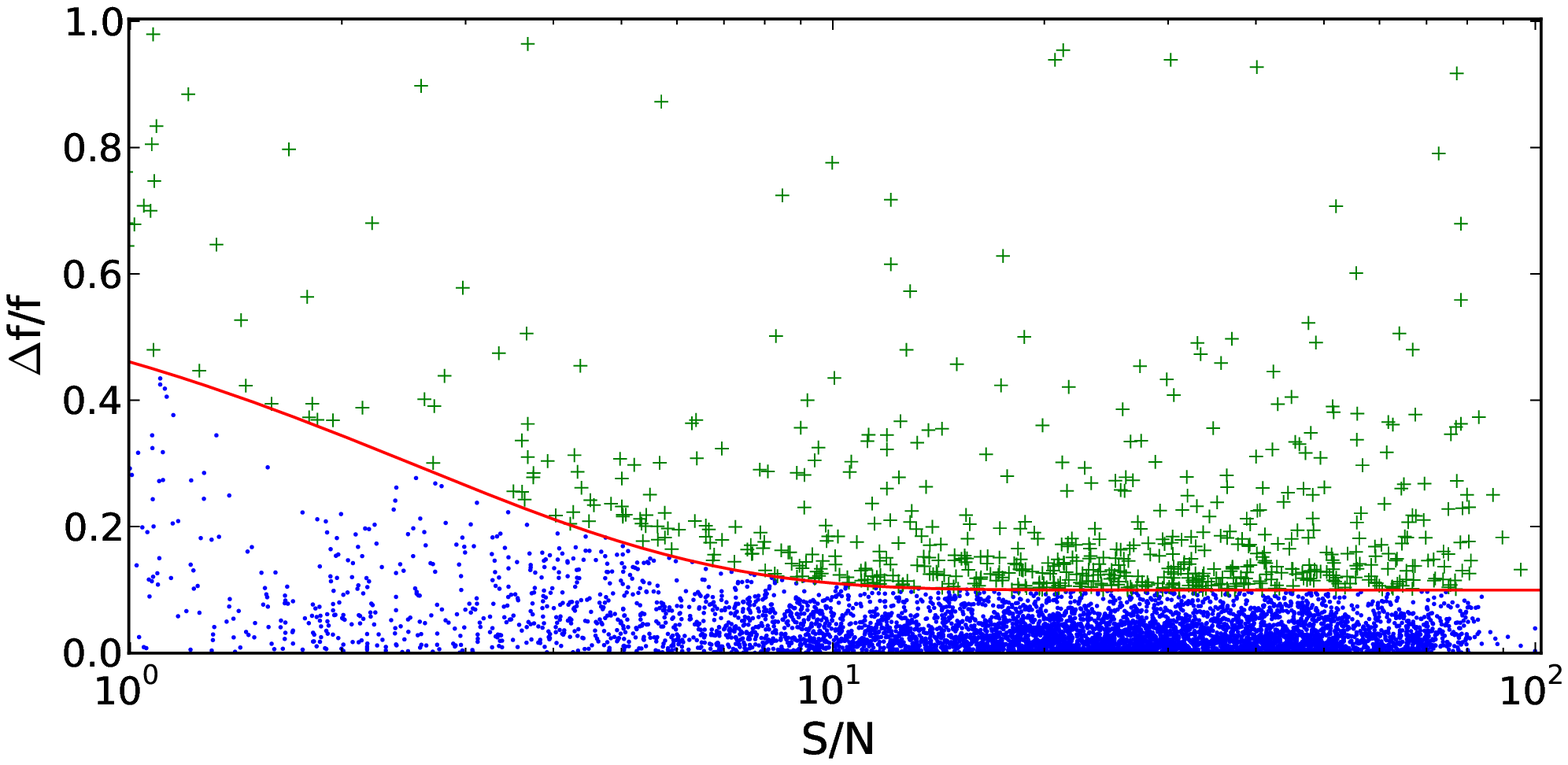} 
\includegraphics[width=0.49\textwidth]{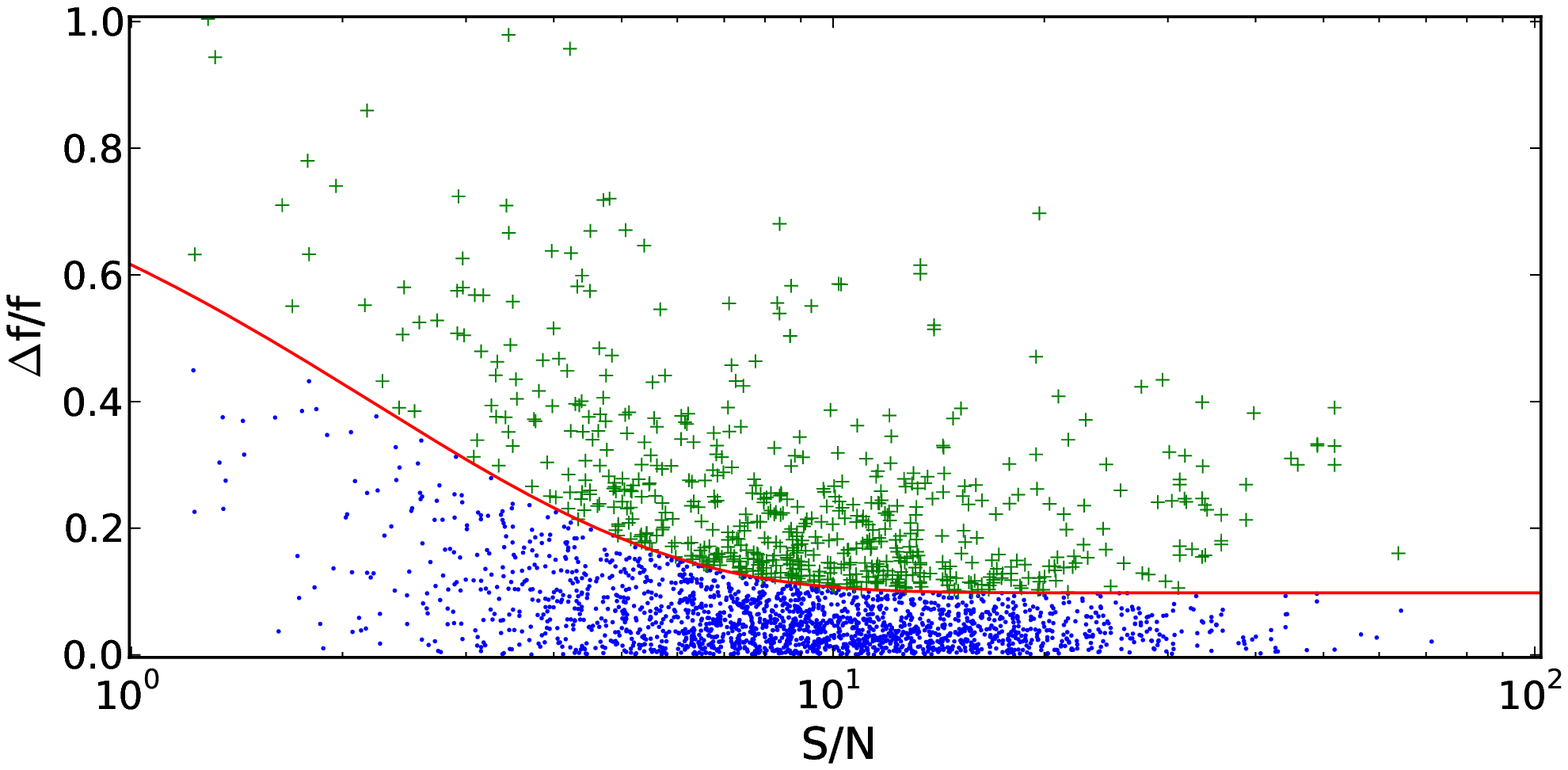} 
\caption{Top Panel: Relative flux change $\Delta f/f$ seen in repeat spectra of stars as a function of the SNR of the high-SNR epoch. The fitted exponential function (red solid line) divides stars deemed to be variable (green crosses) and non-variable (blue points). The non-variable stars are used to calculate a wavelength-dependent spectrophotometric recalibration for pairs of repeat spectra on each plate. Bottom Panel: $\Delta f/f$ as a function of SNR, similar to the top panel, but for known quasars after applying our spectrophotometric recalibration. The sample of variable quasars defined here is used to produce the composite spectra in Figure 5.}
\end{center}
\end{figure}

	For each pair of non-variable stellar repeat spectra, we take the ratio of the lower-SNR 
epoch spectrum to the higher-SNR epoch spectrum; for a non-variable star for which the two 
epochs of spectra are perfectly calibrated, this results in a flat ratio spectrum with ratio 1. However,
plate-wide wavelength-dependent systematic calibration differences between the two epochs 
may be present. We take the median ratio spectrum of all non-variable stars on each plate and
interpolate a 5th-order polynomial to reduce the effects of noise. Prior to the interpolation, we clip 
the top and bottom 3 percentile of pixels in the ratio spectrum to avoid skewing the interpolation 
from outlying pixels. The low-SNR epoch spectra of all quasars on each plate will be multiplied 
by this interpolated median ratio spectrum to match the calibration of the high-SNR epoch spectra. 
The interpolated median ratio spectra used in the spectrophotometric recalibration are generally
a $<$5\% correction at all wavelengths, and are almost all $<$10\%, consistent with the findings of WI05.

	To select all quasars in these 71 pairs of repeat plate observations, we match all spectra to 
the SDSS DR7 quasar catalog \citep{schneider10} to find a total of 2,626 quasars, and apply the 
wavelength-dependent spectrophotometric recalibration for all quasars on each plate. Comparison 
of repeat spectra for objects that did not significantly vary between the two epochs will be 
dominated by noise, and so we place a SNR-dependent variability cut on the $\Delta f/f$ for each 
quasar, similar to the stars, but now to select variable quasars. After binning the quasars into 
13 equal bins of SNR, we fit an exponential to the 75th percentile $\Delta f/f$ in each
bin of the form $\Delta f/f$ = 0.81exp(SNR/$-0.22)+0.10$. We have used the 75th percentile
$\Delta f/f$ in the variability cut rather than the 90th percentile used in the stellar case because quasars 
are known to be more strongly variable than stars in general \citep{se07}. Out of 2,626 quasars 
on the 71 plates, 626 are selected as spectroscopic variables. We note that by design, this sample
of quasars we use to study spectral variability are those exhibiting the strongest variability; this
is desirable for the present study to ensure high SNR of the spectral variability results. 

\begin{figure*}[t]
\includegraphics[width=0.49\textwidth]{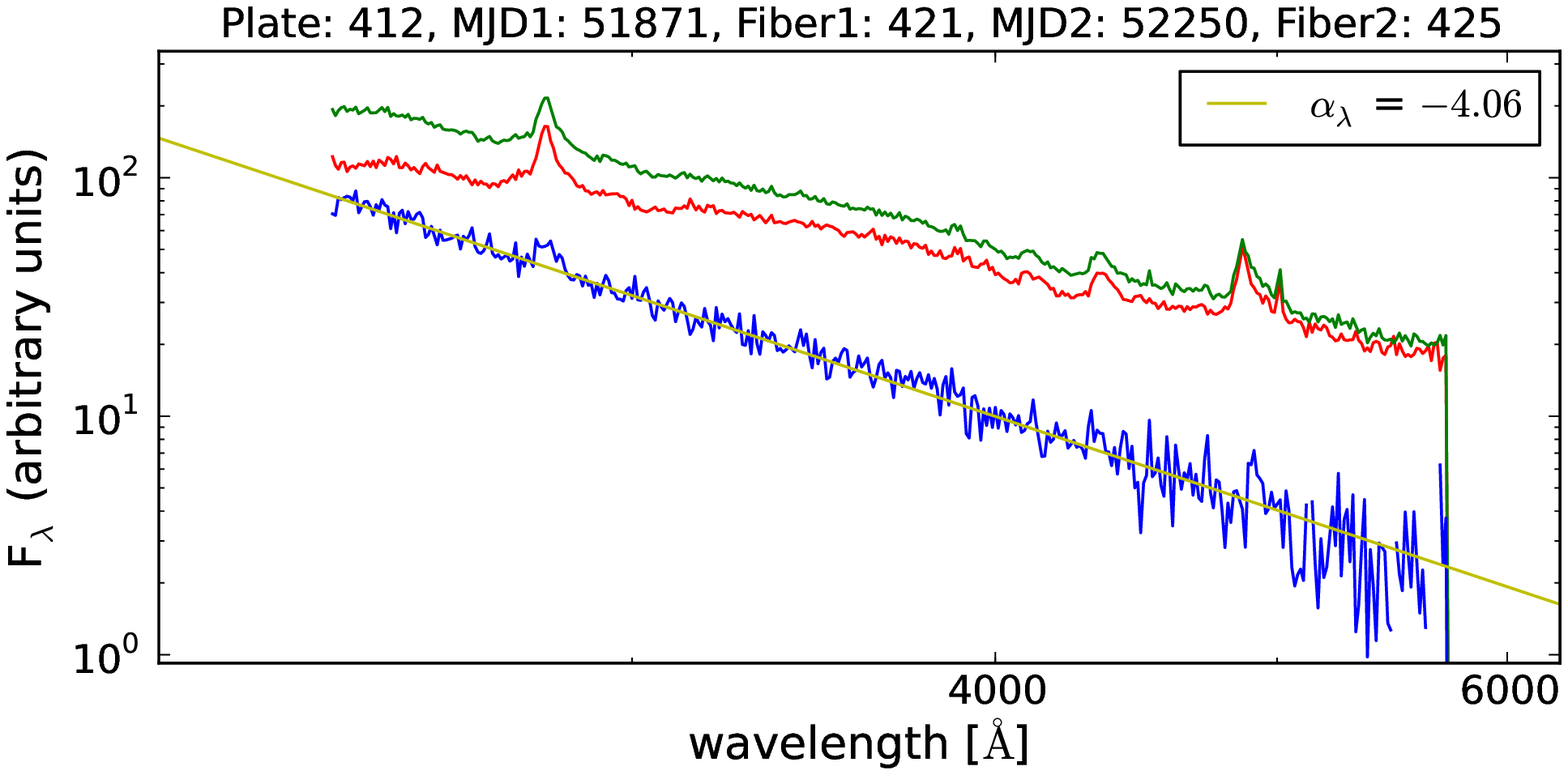}
\includegraphics[width=0.49\textwidth]{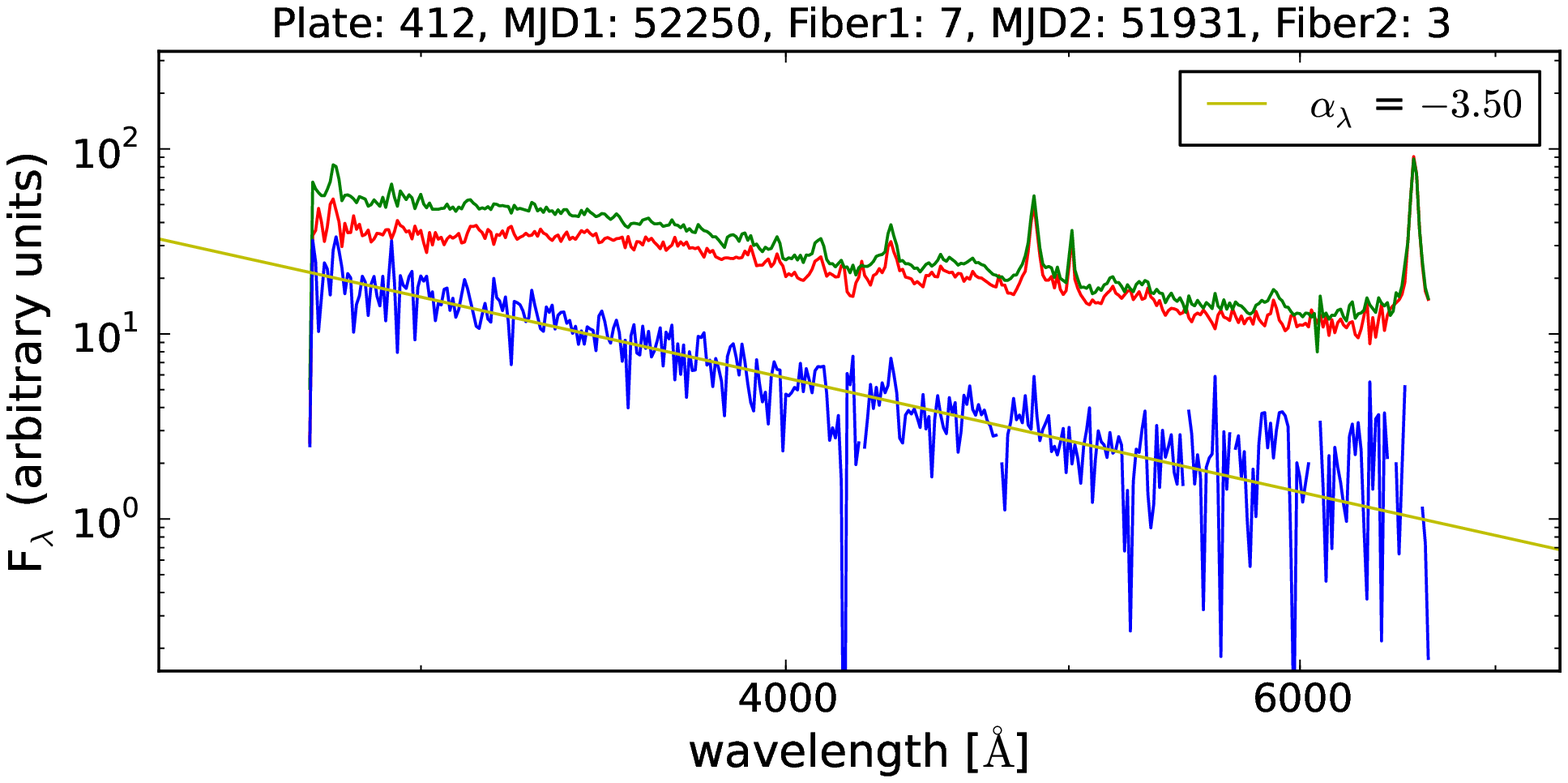}
\includegraphics[width=0.49\textwidth]{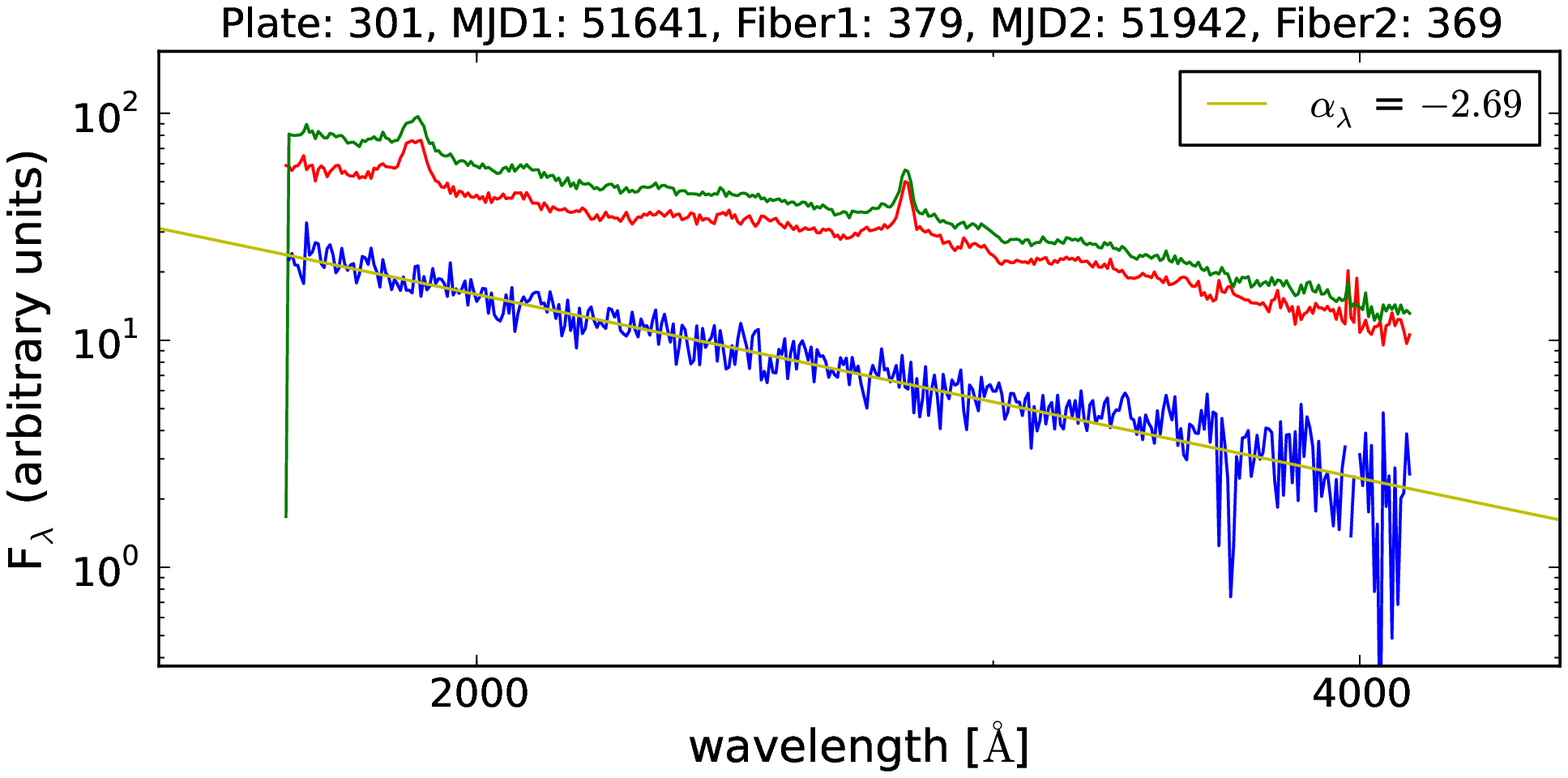}
\includegraphics[width=0.49\textwidth]{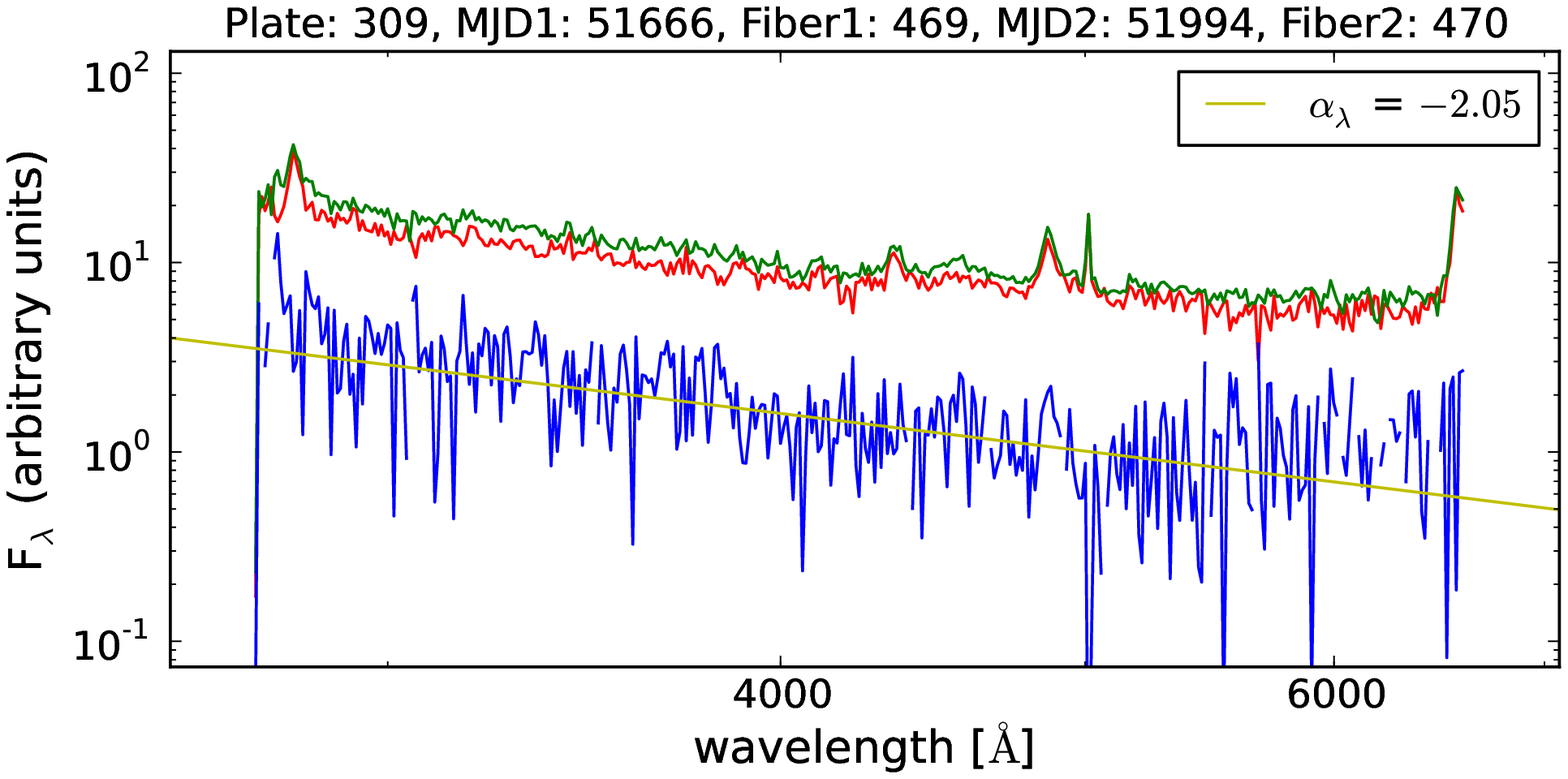}
\includegraphics[width=0.49\textwidth]{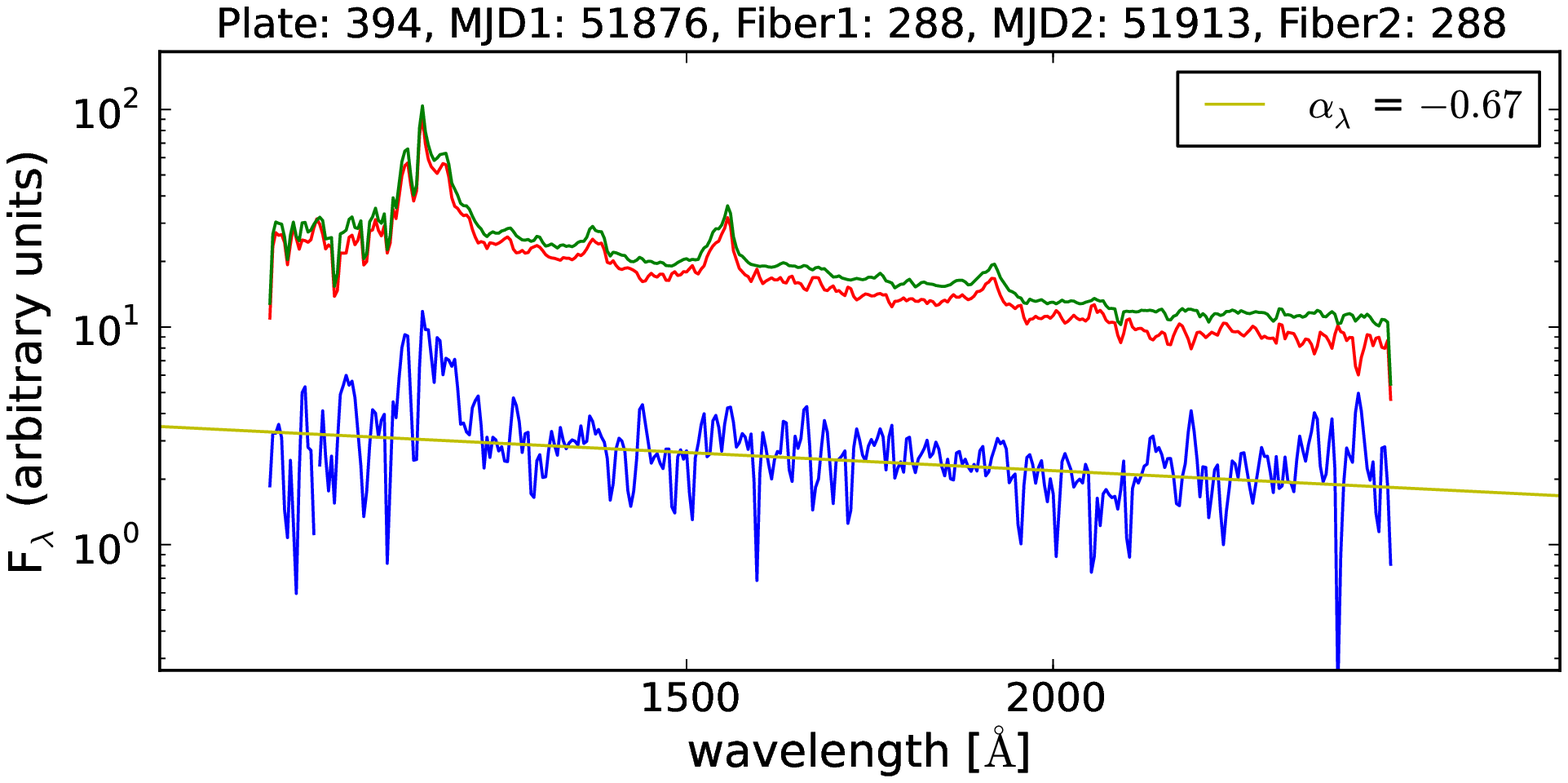}
\includegraphics[width=0.49\textwidth]{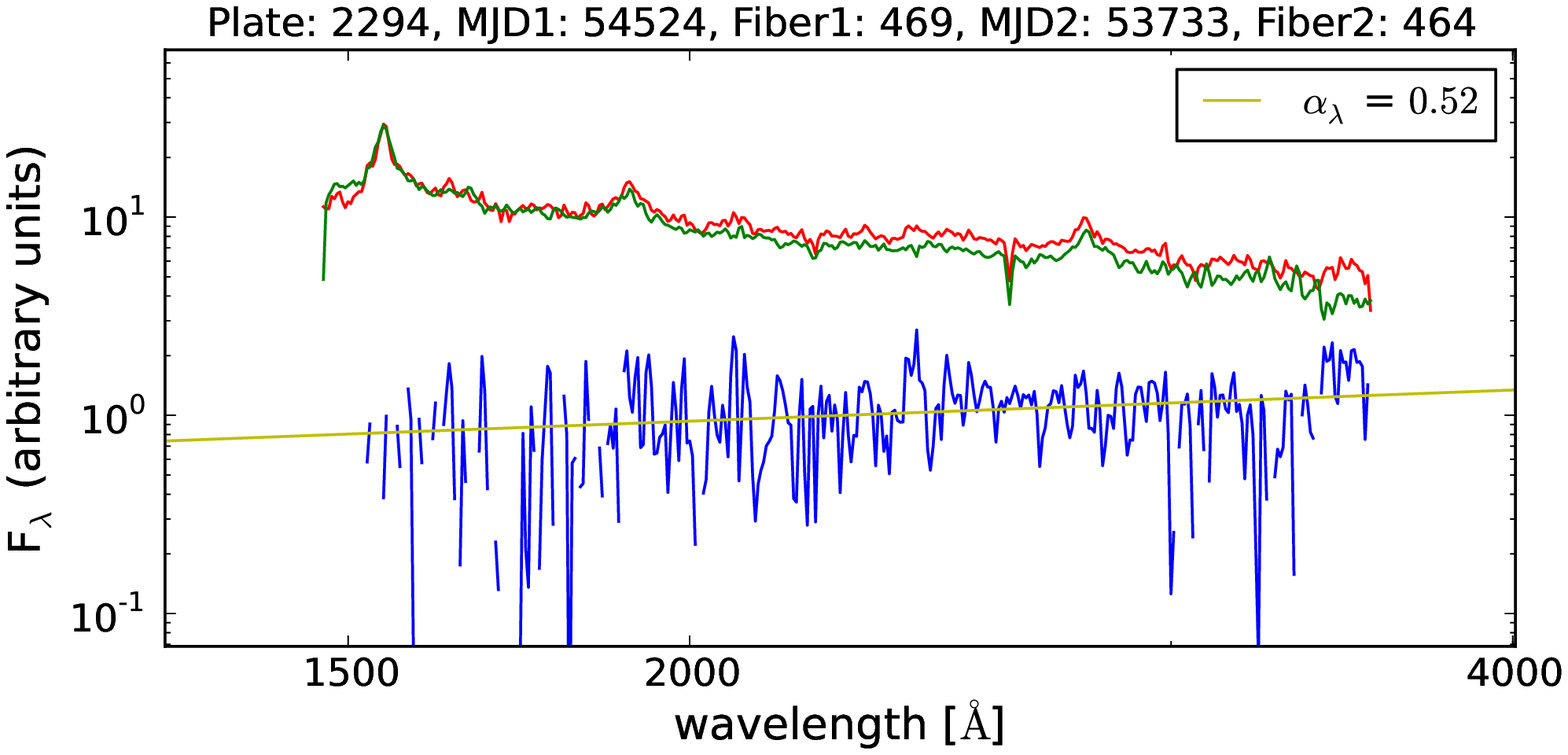} 
\caption{Examples of pairs of repeat variable quasar spectra (green and red), their difference spectra (blue), and best-fit power-laws to the difference spectra continuum (yellow), for a range of power-law indices of the difference spectrum. All spectra shown are corrected for Galactic extinction, and are scaled to arbitrary flux densities. The vast majority of the quasar repeat spectra in our sample show a strong bluer-when-brighter trend.}
\end{figure*}

\section{Quasar Difference Spectra}
\subsection{Difference Spectra and Their Properties}
	Before construction of difference spectra using the 626 pairs of variable quasar spectra 
selected in Section 2, we shift each spectrum to the rest-frame using visually-inspected redshifts 
from \citet{schneider10}. To ensure that the resulting difference spectra are `positive', we subtract
the spectral epoch with the lower integrated flux from the higher, with uncertainties added 
in quadrature. The continuum of each difference spectrum should thus be bluer than either of 
the individual epochs if these quasars exhibit a bluer-when-brighter trend. We visually inspect
all 626 pairs of spectra along with their difference spectra, and find that the continuum of 
the difference spectra are well-fit by power-laws, and indeed show the bluer-when-brighter trend 
in the vast majority of cases. In the visual inspection, we identify 11 pairs of spectra which show 
evidence for strong broad absorption lines (BALs). BAL quasars are known to have atypical continuum 
properties \citep[e.g.,][]{re03, gi09}, and the BALs are know to exhibit intrinsic variability 
in their absorption line strengths over long timescales \citep[e.g., ][]{gi08, gi10, ca11, ca12, fi12, fi13}. 
To avoid contamination,  we remove these 11 BAL quasars from our sample. We also remove 11 
additional quasars identified in the visual inspection for which the variability was clearly dominated 
by noise. The remaining sample of 604 quasar is the sample for which all further results from our 
analysis are reported. 

	We fit the continuum of each difference spectrum to a power-law using a simple $\chi^2$ fit, 
incorporating the uncertainties in the difference spectra. Although broad emission lines are 
well-known to be less variable than the continuum (WI05), there is still evidence of emission line 
variability in our difference spectra. Thus, we mask out the following wavelength regimes dominated 
by broad emission lines in the continuum fitting: 1360-1446\AA~(Si IV, O IV]), 1494-1620\AA~(C IV), 
1830-1976\AA~(C III], Fe III), 2686-2913\AA~(Mg II), 4285-4412\AA~(H$\gamma$), 
4435-4762\AA~(Fe II), 4760-4980\AA~(H$\beta$), 4945-4972\AA~([O III]), 4982-5035\AA~([O III]), 
and 5100-5477\AA~(Fe II), as well as wavelengths $<$1300\AA~to avoid Ly-$\alpha$ 
emission and absorption. The choice of these masked regions are informed by the composite 
SDSS quasar spectrum of \citet{va04}; the numerous other lines and line complexes present in quasar 
optical spectra which we do not mask out tend to be less prominent, and we do not find evidence that 
these other emission lines significantly affect the continuum fitting in our visual inspections.
We clip the top and bottom 1 percentile of pixels in each spectrum after applying these mask 
(but before the fitting) to avoid strong outliers. After the first power-law fit, we again clip the top 
and bottom 1 percentile of pixels away from the best-fit power-law, before refitting the final time. 
	
	Figure 3 shows a few examples of pairs of quasar spectra and their difference spectra from
our sample, for a range in fitted power-law spectral indices $\alpha_\lambda$ (where $F_\lambda$ 
$\propto$ $\lambda^{\alpha_\lambda}$) of the difference spectra continua. The difference spectra
show excellent fits to a simple power-law, and show a strong bluer-when-brighter trend. Figure 4 
shows the distribution of the 
spectral indices of the difference spectra for all 604 quasars, with and without corrections for Galactic 
extinction using the extinction maps of \citet{sc98} and the reddening law of \citet{ca89}. To compare the 
difference spectra to the underlying spectra, we also calculate power-law spectral indices of the 
continua from the high-SN epoch in each pair of repeat quasar spectra; this provides a relatively 
unbiased view of the general spectral properties of each quasar. The fitting of the continua of the 
spectra to a power-law is performed similar to the difference spectra, but with an additional 
wavelength region mask of all wavelengths $>$5800\AA~ to avoid contamination from host-galaxy 
emission \citep{va04}. As expected, Figure 4 shows that quasar difference spectra are generally 
bluer than single-epoch quasar spectra, and the addition of corrections for Galactic extinction causes 
the continuum spectral indices to become even bluer. 

\begin{figure}[t]
\begin{center}
\includegraphics[width=0.49\textwidth]{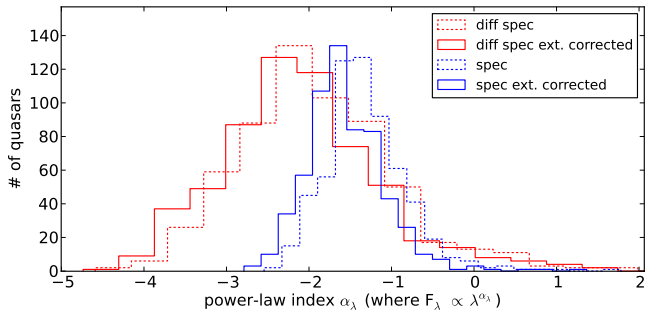} 
\caption{Distribution of fitted power-law spectral indicies $\alpha_\lambda$ for the high-SN epoch spectra of all variable quasars in our sample (blue dotted), and for their difference spectra (red dotted) without correcting for Galactic extinction. The distributions of the fitted indices for the same quasars after correcting for Galactic extinction are shown in solid. The median of these distributions are an excellent match to previous studies. 
}
\end{center}
\end{figure}

\subsection{Composite Spectrum and Difference Spectrum}
	We construct a geometric mean composite spectrum for both the difference spectra as well 
as the high-SNR single-epoch spectra of the 602 quasar in our sample with corrections for Galactic 
extinction; the use of a geometric mean to construct the composite spectrum preserves the 
arithmetic mean power-law spectral indices and extinction of a sample of power-law spectra 
\citep{re03}, well-suited for our investigation of quasar continuum properties. The composite 
spectra are created by using our power-law fits to the high-SNR epoch and difference spectra for 
each pair of spectra, and scaling the flux density of the fitted power-laws at 3062\AA~(a relatively 
line-free wavelength and covered by nearly all of the spectra in our sample) to a fixed arbitrary 
flux density. The spectra and difference spectra themselves are then each rescaled by the same 
scaling factor as their fitted power-laws, and a geometric mean of all rescaled spectra and difference 
spectra in each wavelength bin is taken to produce the composites. The geometric mean composite 
spectrum and composite difference spectrum are shown in Figure 5. The 1$\sigma$ uncertainties on 
the composite spectrum and difference spectrum is estimated by resampling all the pixel flux densities 
in each spectrum from a Gaussian with center at the measured flux density and width set to the 
uncertainty in the flux density. All 604 pairs of spectra are resampled 10$^3$ times, and 10$^3$ 
composite spectra and composite difference spectra are produced. The 1$\sigma$ uncertainties shown
are the 1$\sigma$ spreads in these 10$^3$ resampled composites around the mean.

	We fit the continuum of the composite spectrum and composite difference spectrum using the 
same broad emission line wavelength masks as before to find a power-law spectral index of $\alpha_{\lambda,comp} = -1.56 \pm 0.01$ for the composite spectrum, and $\alpha_{\lambda,diff} = 
-2.12 \pm 0.02$ for the composite difference spectrum. Without corrections to each spectrum for 
Galactic extinction (not shown), the spectral indices are $\alpha_{\lambda,comp} = -1.38 \pm 0.01$ 
and $\alpha_{\lambda,diff} = -1.94 \pm 0.02$. We note that although the host-galaxy emission that 
dominates the composite quasar spectrum redward of $\sim$6000\AA~should not be time-variable, 
the composite difference spectrum seems to show some host-galaxy emission residuals. This is 
predominately due to noisy data and poorly-subtracted masked pixels at the red edge of the spectra, 
as well as the fact that the number of spectra contributing to the composite at wavelengths 
$>$6000\AA~is only $\sim$50 out of a total sample of 604. The composite difference spectrum 
is thus unreliable at $>$6000\AA.

\section{Comparison to Previous Studies}
	Our analysis thus far has closely followed the work of WI05, and in this section we 
compare our results to those from previous studies. For the composite quasar spectrum 
continuum, our calculated spectral index $\alpha_{\lambda,comp} = -1.56 \pm 0.01$ with 
corrections for Galactic extinction is an exact match to the results of \citet{va04}, while our 
$\alpha_{\lambda,comp} = -1.38 \pm 0.01$ without corrections for Galactic extinction is an 
excellent match to the $\alpha_{\lambda,comp} = -1.35$ calculated by WI05, all based on 
SDSS-I/II spectra in a similar range of wavelengths. For the composite quasar difference 
spectrum continuum, our calculated $\alpha_{\lambda,diff} = -1.94 \pm 0.02$ without 
corrections for Galactic extinction is also an excellent match to the $\alpha_{\lambda,diff} 
= -2.00$ from WI05. The minor discrepancy is likely due to differences in sample size, as 
our sample of 602 quasars is approximately a factor of two larger than the sample used in 
WI05 (although of course many objects are in common).
	
	\citet{pe06} utilized the composite difference spectrum generated by WI05 and showed 
that it is well-fit by synthetic difference spectra generated from a thin disk model which has 
undergone some change in its global accretion rate. However, difference spectra are subject
to both Galactic and intrinsic (host-galaxy) extinction; the composite difference spectrum of WI05
was not corrected for any extinction, and thus should not be directly compared to models.
Although we have corrected each spectrum for Galactic extinction in our work, the intrinsic 
extinction of each quasar is much more difficult to take into account. Thus, our composite
difference spectrum is subject to unknown amounts of intrinsic extinction from each of the 
individual quasars in the sample.

\begin{figure}[t]
\begin{center}
\includegraphics[width=.49\textwidth]{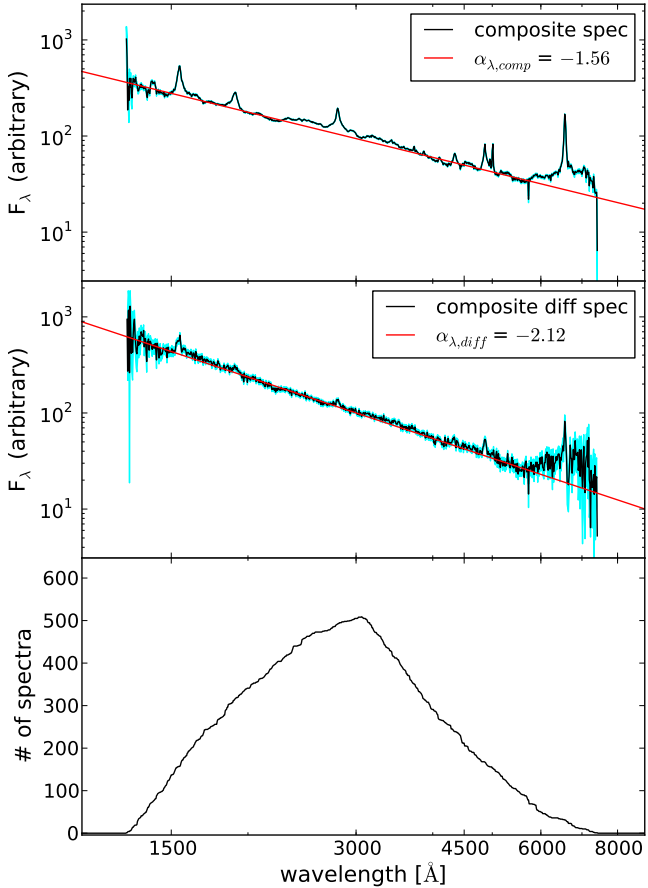} 
\caption{Top Panel: Geometric mean composite quasar spectrum of all high-SNR epoch spectra in our sample of 602 variable quasars, corrected for Galactic extinction, with 1$\sigma$ uncertainties shown in light blue. The fitted power-law index of the continuum ($\alpha_{\lambda,comp} = -1.56$) is in excellent agreement with the composite SDSS quasar spectrum of \citet{va04} for $\lambda$ $\textless$ 5000 \AA, where host-galaxy light is minimal. Middle Panel: Geometric mean composite quasar difference spectrum for the pairs of variable quasars in our sample, corrected for Galactic extinction, also with 1$\sigma$ uncertainties shown in light blue. The fitted power-law index of the continuum ($\alpha_{\lambda,diff} = -2.12$) is steeper than that of the composite spectrum, showing that the majority of quasars exhibit a bluer-when brighter trend. Bottom Panel: Number of spectra contributing to the composite spectrum and composite difference spectrum at each wavelength.}
\end{center}
\end{figure}

\begin{figure}[t]
\begin{center}
\includegraphics[width=.49\textwidth]{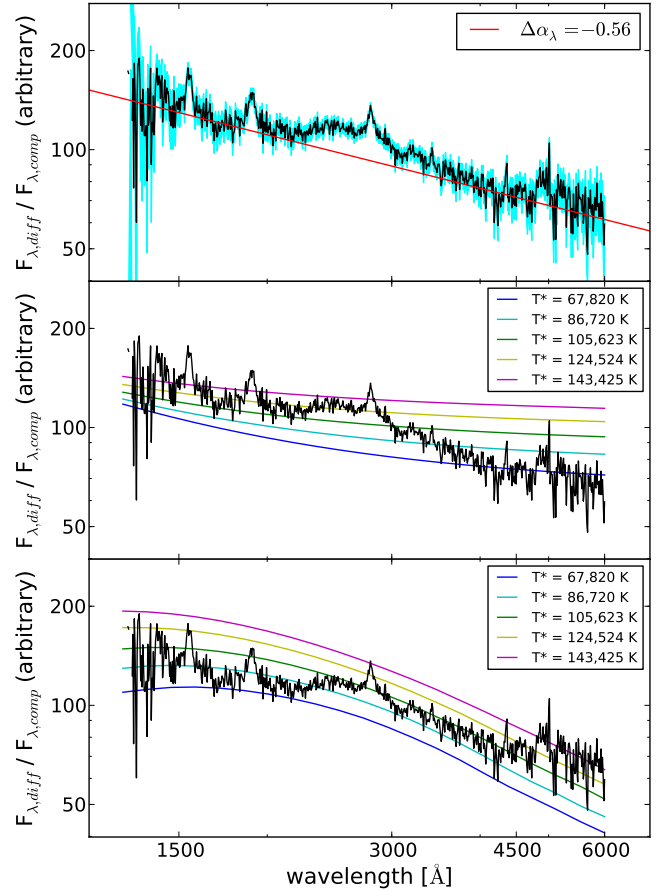} 
\caption{Top Panel: Observed composite relative variability spectrum, created by dividing the composite difference spectrum by the composite spectrum from Figure 5, with 1$\sigma$ uncertainties shown in light blue. The best fit power-law to the continuum is shown (red), with relative spectral index $\Delta\alpha_\lambda$ $\equiv$ $\alpha_{\lambda,diff}$ $-$ $\alpha_{\lambda,comp}$ = $-0.56$. We note that we have flipped this spectrum in the vertical direction around the best-fit power-law continuum (see text for details). Middle Panel: Comparison of the relative variability spectrum to that generated from changing the accretion rate in a thin disk model by 5\%. The resulting synthetic relative variability spectra for a range of thin disk characteristic temperatures are shown; none are as steep as the observations, and thus do not display as strong of a bluer-when-brighter trend. Bottom Panel: Same as the middle panel, but instead comparing synthetic relative variability spectra generated from an inhomogeneous accretion disk with large temperature fluctuations. This model provides a good match to our observations in the optical.
}
\end{center}
\end{figure}

	To avoid issues with extinction and robustly compare our observed spectral variability
to models, we instead consider the spectral variability \emph{relative} to the underlying spectra 
of the quasars, by dividing the geometric mean composite difference spectrum by the geometric
mean composite spectrum from Figure 5. The result, which we call the `relative variability 
spectrum', is shown in Figure 6 and has power-law spectral index of $\Delta\alpha_\lambda$ 
$\equiv$ $\alpha_{\lambda,diff}$ $-$ $\alpha_{\lambda,comp}$ = $-0.56 \pm 0.02$. The result of this
division of geometric mean composite spectra is independent of any extinction with the reasonable 
assumption that the extinction does not significantly change between each pair of observations. 
We note that because broad emission lines are less variable than the continuum, the relative 
variability spectrum will have inverted emission lines; we have flipped the relative variability 
spectrum in Figure 6 in the vertical direction, centered on the best-fit continuum to aid in 
identifying continuum and emission features visually. The spectral variability of quasars relative 
to their underlying spectra was also investigated in WI05 using the ratio between composite 
difference spectra to composite spectra of their quasar sample, leading WI05 to conclude that 
quasars exhibited spectral variability only at wavelengths $<$2500\AA. However, this was done 
by WI05 using arithmetic mean composites, which do not preserve the mean power-law indices 
of the spectra (making interpretation difficult), and is subject to the effects of extinction. In contrast 
to WI05, our relative variability spectrum avoids both these issues by using geometric mean 
composite spectra, facilitating robust comparison to models.

	In Figure 6 (middle panel), we compare our observed  relative variability spectrum to 
synthetic relative variability spectra generated from thin disk models in which the global 
accretion rate has increased by 5\%, for a range in characteristic disk temperatures
\begin{equation}
\label{equ_tstar}
  T^*
\equiv
  \left\{ {3 \dot M G M_{BH} \over 8 \pi r_i^3 \sigma_s}
  \right\}^{1/4}
\end{equation}
\citep{fr02}. The range in $T^*$ in Figure 6 (middle panel) is chosen to span the full range generated 
for thin disks with log$_{10}L/L_{\rm{Edd}} = [-1.1, -0.8]$, and log$_{10}M_{\rm{BH}} = [8.5, 9.5]$.
These ranges in log$_{10}L/L_{\rm{Edd}}$ and log$_{10}M_{\rm{BH}}$ are representative of these 
values in our sample of 604 quasars, which have median log$_{10}L/L_{\rm{Edd}} = -0.89$ and 
median log$_{10}M_{\rm{BH}}  = 8.83$ M$_\odot$ from the catalog of \citet{sh11}. From the spectral variability
shown in Figure 6 (middle panel), it is clear that a scenario in which a thin disk changes its global 
accretion rate cannot account for the strong bluer-when-brighter trend observed in quasars, and it 
does not produce our observed power-law relative variability spectrum. Our spectral variability 
evidence against global accretion rate fluctuations as the cause of quasar flux variability is independently 
in agreement with the argument that quasar accretion disks are too large to vary coherently.
We note that although we have shown in Figure 6 (middle panel) difference spectra of a thin 
disk in which the accretion rate changed by 5\%, the shapes of difference spectra from thin disk 
models with accretion rate fluctuations are not particularly sensitive to the exact change in accretion 
rate (i.e. the fit will not significantly improve by increasing the change in accretion rate). 
Instead, the difference spectra from thin disk models depend mainly on the disk's characteristic 
temperature, for which we have shown a wide range in Figure 6 (middle panel). A similar 
conclusion was reached by \citet{pe06}.

	Aside from our results, the photometric quasar spectral variability study of SC12
also argued against global accretion rate changes as the sole driver in quasar spectral
variability. We suggest that the source of the discrepancy between the conclusions of SC12 
and \citet{pe06} may be due to their subtly different parameterizations of quasar spectral variability 
and its effects on the extinction. SC12 fitted linear relations to multi-epoch photometry of quasars 
in several filters in magnitude-magnitude space, and transformed the relation into color-magnitude 
space to investigate the variability in different filters. The spectral variability parameter that SC12 
compared to models was the slope of the fit in color-magnitude space after the transformation;
this slope was fit after all magnitudes for each quasar were normalized to its mean 
magnitudes for the different filters (see Equation 3 in SC12), and thus parameterizes the 
spectral variability of each quasar \emph{relative} to its underlying color. This parameterization 
is similar to the power-law spectral index of the relative variability spectrum ($\Delta\alpha_\lambda$) 
we calculate, but for photometric colors, and is thus also independent of extinction. Although this 
subtlety was not discussed in SC12, this may be the source of the conflicting results between SC12 
and \citet{pe06} in comparing their observations to global accretion rate fluctuations in a thin disk. 

	Aside from a simple thin disk, SC12 also compared their spectral variability results to more 
sophisticated models presented in \citet{da07}, finding that their data cannot be explained 
by accretion rate changes in any of these disk models. SC12 suggested that ephemeral hotspots 
may be needed to match their observations; in the next section, we compare our observed results to 
one such time-dependent model of a simple inhomogeneous disk.  

\section{Disk Models with Localized Temperature Fluctuations}

	\citet{de11} presented a simple analytic model of a time-dependent inhomogeneous disk, 
based on a thin disk radiating with independent zones undergoing temperature fluctuations 
and emitting locally as a blackbody. Aside from predictions of MHD turbulence in simulations 
of accretion disks, there is now also observational evidence for disks with time-dependent 
temperature fluctuations from microlensing disk-size measurements, the strong UV spectral
continuum, and single-band variability characteristics. \citet{de11} find that to satisfy these 
observational constrains, accretion disks must be $strongly$ inhomogeneous, with large 
localized temperature fluctuations. These large temperature fluctuations in the disk 
inhomogeneities will likely cause the spectrum to be highly variable at short wavelengths, 
and produce distinct spectral variability with which we will compare our observations.

	We set up the inhomogeneous accretion disk model of \citet{de11}, starting with a standard thin 
disk with an inner edge at the innermost stable circular orbit of a non-spinning black hole,
and dividing its surface into $n$ zones per factor of two in radius. The zones are roughly equally 
divided radially and azimuthally, although the exact setup does not noticeably affect the results in 
our tests. The logarithmic temperature log$_{10}T$ in each zone independently fluctuates as a 
first-order continuous autoregressive (CAR(1)) process, motivated by studies of single-band 
variability characteristics \citep[e.g.,][]{ke09}. The mean temperature in each zone is set to the 
log$_{10}T$ of the thin disk model at that radius, and the constant driving the log$_{10}T$ 
fluctuations in the CAR(1) process is $\sigma_T$. The characteristic decay timescale of the 
temperature fluctuations is set to 200 days in the rest-frame, motivated by the observed timescales
in \citet{ke09} and \citet{ma10}. Our spectral variability results are not sensitive to the choice of this 
timescale, although we note that if the decay timescale is significantly longer than the time-lag 
between repeat observations, the quasar will not appear to be significantly variable. All regions in 
the inhomogeneous disk are assumed to emit locally as a blackbody, and no relativistic effects are 
considered in this simple model.

	We run the inhomogeneous disk model, sampling its spectrum in the wavelength range of 1300-5800\AA~at 50 day intervals in the rest-frame, after an initial `burn in' time of 500 days to 
allow the disk to become inhomogeneous. To faithfully compare to our observations, we calculate 
the change in observable flux between successive time-steps by integrating the spectrum, and 
calculate the difference spectrum between any two successive time-steps in which the total flux 
changed by more than 10\%, similar to the variability cut placed on the observed spectra in Figure 2. 
We run the model until we produce $5\times10^3$ synthetic difference spectra, and produce a 
synthetic geometric mean composite spectrum and composite difference spectrum, by renormalizing 
each spectrum in the same way as our SDSS spectra.

	Figure 7 shows synthetic relative difference spectra generated from the inhomogeneous 
disk model over a range in independent zones $n$ and temperature fluctuations $\sigma_T$,
for the same disk temperatures as shown in Figure 6. The inhomogeneous disk produces 
relative variability spectra with a characteristic shape that is a power-law at optical wavelengths,
which flattens at shorter wavelengths in the UV. For increasing $n$ and $\sigma_T$ in Figure 7, 
the relative variability spectra remains power-law-like further into the UV before flattening
(the physical cause of this is discussed in Section 6.1). Thus, to produce the observed power-law 
spectral variability in Figure 6 (top panel), the inhomogeneous disk needs to have many independent 
zones (large $n$) with large temperature fluctuations (large $\sigma_T$).  

	An example comparison of our observed relative variability spectrum to that from the inhomogeneous 
disk, using parameters $n = 10^{2.7}$ and and $\sigma_T = 0.45$, is shown in Figure 6 (bottom panel).
The large temperature fluctuations and number of fluctuating zones required of inhomogeneous 
disks to provide such good fits to our observations is in excellent agreement with the $n = 10^{2.5-3}$ 
and $\sigma_T = 0.35-0.5$ range found by \citet{de11} to simultaneously satisfy independent 
observational constraints from microlensing disk-size measurements, their strong UV spectral continuum, 
and single-band variability characteristics of quasars. This independent result based on spectral 
variability adds to the mounting evidence for large temperature fluctuations in strongly inhomogeneous 
quasar accretion disks. We note that because the relative variability spectrum is constructed from 
composite spectra and composite difference spectra of many quasars, each with different Eddington 
ratios and disk temperatures, the fit of the inhomogeneous disk relative variability spectrum is likely to 
improve with a more complete consideration of these variations.

\begin{figure*}
\begin{center}
\includegraphics[width=.96\textwidth]{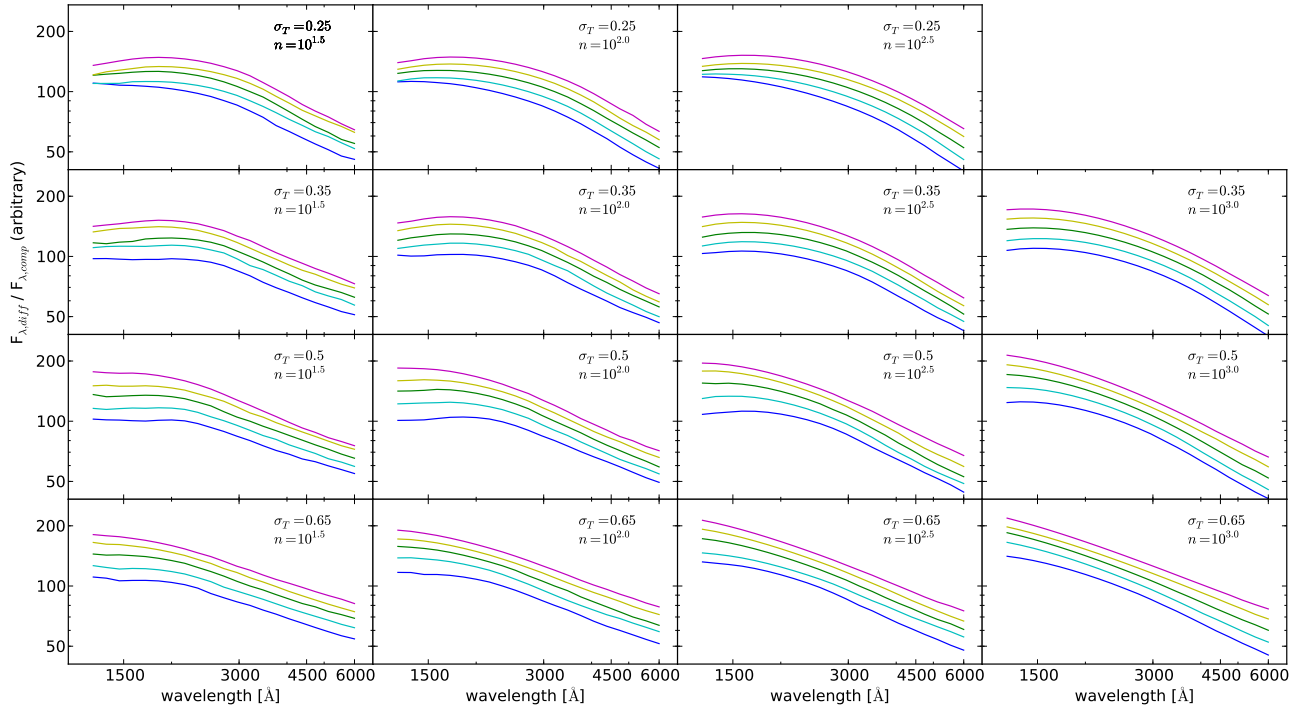} 
\caption{Synthetic relative variability spectra generated form the inhomogeneous disk over 
disk parameters $\sigma_T$ and $n$ (see Section 5 for details). The range in disk temperatures 
shown for each panel is the same as in Figure 6 (middle panel). These synthetic 
relative variability spectra are power-laws in the optical, with a flattening in the UV. The 
turnover occurs further into the UV with both increasing $\sigma_T$ and $n$ (see Section 6.1 
for discussion). Note that we do not show results for larger $n$ and smaller $\sigma_T$ because
disks with these parameters are not variable enough in flux to produce difference spectra
that pass our 10\% flux change cut.
}
\end{center}
\end{figure*}

\section{Discussion}
\subsection{The Connection Between Disk Properties and Resultant Spectral Variability}
	
	Our success of modeling quasar spectral variability
in the previous section can be understood as the result of the confluence of large temperature 
fluctuations in a strongly inhomogeneous disk emitting locally as a blackbody. A flare in a portion 
of the disk causes the variable part of its optical spectrum to be due to the blackbody emission from 
the flaring region. For very high-temperature flares, the blackbody spectrum of the flaring 
region peaks well into the UV, and thus the difference spectrum is dominated by its power-law 
Rayleigh-Jeans tail. This is the cause of the trend in Figure 7 where the power-law portion of the model 
relative variability spectrum increasingly extends into the UV as $\sigma_T$ increases. We note that 
for small $\sigma_T$, the turnover in the UV is a flattening rather than a sharp peak because it 
is the superposition of the blackbody peaks of numerous flaring regions.

	Aside from the large temperature fluctuations, the disk must also be strongly inhomogeneous,
with a large number of independently fluctuating regions (i.e. a disk with large $n$). This is due to
the fact that the total flux variability amplitude of the disk scales as $N^{-1}$, where $N$ is number of independently varying zones \citep{de11}; as $n$ increases and the disk becomes more strongly inhomogeneous, the total flux variability decreases. In the case of a disk with small $n$, the difference 
spectra will be dominated by smaller, lower-temperature flares with blackbody peaks in the optical rather 
than the UV, and thus the spectral variability will not be dominated by the Rayleigh-Jeans spectrum. 
Conversely, for large $n$, the disk does not exhibit strong overall flux variability, and only very large, 
high-temperature flares are actually observable. Thus, the observed flux variability in strongly 
inhomogeneous disks will be dominated by the Rayleigh-Jeans spectrum, naturally producing
power-law spectral variability. This effect causes the power-law portion of the model relative variability 
spectrum in Figure 7 to extend further into the UV with increasing $n$. We emphasize that this line of 
evidence for large temperature fluctuations in strongly inhomogeneous disks is independently in 
excellent agreement with other observational constraints from microlensing disk sizes, 
their strong UV spectral continuum, and single-band variability properties. 

	We note that because the temperatures in each independently fluctuating zone is damped in 
the CAR(1) process and thus cannot increase infinitely, the inhomogeneous disk predicts a flattening 
in the relative variability spectrum in the UV (e.g. as seen in Figure 7). This flattening is not seen in 
our difference spectra, which appears to be well-fit by a power-law down to $\sim$1300\AA, although the 
SNR decreases dramatically below $\sim$1500\AA. Broadband studies of quasar UV variability have 
shown that quasars are indeed generally more variable in the UV than optical \citep{ge13}. However, 
a more careful investigation of quasar spectral variability from optical to the UV will require contemporaneous optical-UV observations, and will be a fruitful test of the inhomogeneous disk model. We also note that in the 
simple inhomogeneous disk model of \citet{de11}, the global accretion rate is assumed to be constant. 
This may not be entirely justified, as processes causing the large temperature fluctuations such as the MRI 
are likely to also induce fluctuations in the accretion rate (e.g. by causing the viscosity to change locally). 
Although we have ruled out accretion rate fluctuations as the sole driver of quasar spectral variability 
in Section 4, the observed difference spectra in Figure 5 (middle panel) may still be affected by
changes in the accretion rate. In particular, because Figure 6 (middle panel) shows that accretion
rate changes produce spectral variability that is particularly strong in the UV, the addition of accretion
rate changes to the inhomogeneous disk model may cause difference spectra to become power-law
like well into the far UV, further improving the match to the observed UV spectral variability. 

\subsection{Other Possibilities for Difference Spectra}
	The disk models we have considered all assume that the disk emits locally as a blackbody;
while this may be a good approximation for the underlying disk spectrum, it is not entirely 
justified for difference spectra, which instead isolate the variable part of the spectrum. 
It is possible that the observed power-law difference spectrum is instead at least partially due to 
non-thermal emission, which could dominate the variable portion of quasar optical/UV spectra. For 
example, disk inhomogeneities from the photon bubble instability in the magnetized atmosphere 
of a radiation pressure-dominated disk \citep{ar92, ga98, tu05} may cause the spectrum during 
flaring epochs to become non-thermal, due the shorter paths in the low gas-density bubbles for 
photons to diffuse to the photosphere \citep{ga98}, or due to an increase in free-free emission in 
the high photon-density bubbles \citep{da09}. Radio-loud quasars have long been suspected to 
harbor weak or unresolved jets, and the highly-variable jet synchrotron emission can also produce 
power-law difference spectra in the optical/UV. Future observations of quasar spectral variability
across multiple wavelength regimes contemporaneously can help constrain these possibilities. 

\subsection{Variability Correlations with M$_{BH}$ and Luminosity}
	Finally, we note that our findings that quasar spectral variability cannot be driven purely 
by global accretion rate changes in a thin disk is not at odds with the trends between variability
amplitude, black hole mass, and luminosity that many studies have found, and which suggest that 
the Eddington ratio may be driving these trends. It is still possible that differences in the mean 
Eddington ratio among different quasars drive their variability properties, a conclusion also 
reached by SC12. Notably, \citet{ma10} also found that the scaling relation between these 
quantities is much shallower than that predicted from the Eddington ratio, suggesting additional 
physics may be necessary to explain the observed trend. Modeling of these trends using 
inhomogeneous disk models awaits future investigation.

\section{Conclusions}

	The characteristic flux variability of quasars reflects complex processes in the accretion 
disk, yet the cause of the variability is still unknown. Aside from single-band variability 
characteristics, the spectral variability of quasars provides an additional, independent constraint 
on models of the variability mechanism. Using repeat spectra of quasars in SDSS-I/II, we 
investigate the optical spectral variability of quasars, which are known to show a 
bluer-when-brighter trend. After a wavelength-dependent spectrophotometric recalibration of the 
quasar spectra using non-variable stars observed on the same plates, we construct difference 
spectra of 602 variable quasars in our sample, thus isolating the variable part of the spectrum. 
We compare our observations to synthetic difference spectra generated from a thin disk model 
in which the accretion rate has varied, as well as a simple inhomogeneous disk model with 
localized temperature fluctuations. In particular, we find the following:

\begin{enumerate}
\item Quasar difference spectra appear to be power-laws with spectral indices steeper than 
their single-epoch spectra, indicating that the vast majority of quasars show a bluer-when-brighter
trend. We measure quasar spectral variability using the relative variability spectrum, which is 
independent of any extinction. We find that accretion rate fluctuations in a thin disk model cannot 
produce the strong bluer-when-brighter trend observed. This is contrary to the results of some 
previous investigations, and may be due to the effects of intrinsic and Galactic extinction that were 
not accounted for in those previous studies. 

\item A time-dependent inhomogeneous disk model can produce spectral variability that provides 
a good match to our observations over optical wavelengths if the disk is strongly inhomogeneous,
with large temperature fluctuations. The difference spectra produced by such inhomogeneous disks 
are dominated by the Rayleigh-Jeans spectrum from the hot `flaring' regions in the optical, and are thus 
naturally power-laws. The large temperature fluctuations and large number of zones in the disk 
required to match our observed spectral variability is in excellent agreement with independent 
observational constraints from quasar microlensing disk-sizes, their strong UV spectral continuum, 
and single-band flux variability characteristics. 
\end{enumerate}

Our spectral variability constraints suggests that future quasar disk models should be time-dependent, 
and include large temperature fluctuations. Improved global GRMHD simulations of radiative disks will 
help inform more sophisticated inhomogeneous disk models to compare to observations. Observationally, 
future time-domain photometric and spectroscopic observations from the Large Synoptic Survey 
Telescope \citep{iv08} and the Time-Domain Spectroscopic Survey portion of SDSS-IV will help better 
constrain these models.

\acknowledgments
	JJR thanks Yusra AlSayyad and James R.~A.~Davenport for helpful discussions
about processing SDSS spectra. Support for JJR was provided by NASA through Chandra 
Award Numbers AR9-0015X, AR0-11014X, and AR2-13007X, issued by the Chandra X-ray Observatory 
Center, which is operated by the Smithsonian Astrophysical Observatory for and on behalf of
NASA under contract NAS8-03060.
Support for EA was provided by NASA through a grant from
the Space Telescope Science Institute, which is operated by the Association of Universities for 
Research in Astronomy, Inc., under NASA contract NAS 5-26555
	
	Funding for the SDSS and SDSS-II has been provided by the Alfred P. Sloan Foundation, 
the Participating Institutions, the National Science Foundation, the U.S. Department of Energy, 
the National Aeronautics and Space Administration, the Japanese Monbukagakusho, the Max 
Planck Society, and the Higher Education Funding Council for England. The SDSS Web Site is 
http://www.sdss.org/.
	
	The SDSS is managed by the Astrophysical Research Consortium for the Participating 
Institutions. The Participating Institutions are the American Museum of Natural History, 
Astrophysical Institute Potsdam, University of Basel, University of Cambridge, Case Western 
Reserve University, University of Chicago, Drexel University, Fermilab, the Institute for 
Advanced Study, the Japan Participation Group, Johns Hopkins University, the Joint Institute
for Nuclear Astrophysics, the Kavli Institute for Particle Astrophysics and Cosmology, the Korean 
Scientist Group, the Chinese Academy of Sciences (LAMOST), Los Alamos National Laboratory, 
the Max-Planck-Institute for Astronomy (MPIA), the Max-Planck-Institute for Astrophysics (MPA), 
New Mexico State University, Ohio State University, University of Pittsburgh, University of Portsmouth, 
Princeton University, the United States Naval Observatory, and the University of Washington.

\bibliography{bibref}
\bibliographystyle{apj}

\end{document}